\documentclass[fleqn,usenatbib]{mnras}
\usepackage{newtxtext,newtxmath}
\usepackage{graphicx} 
\usepackage{caption}
\usepackage{subcaption}
\usepackage[T1]{fontenc}
\usepackage{float}
\usepackage{cleveref}

\newcommand{\msun}{\,\mathrm{M}_\odot}

\newcommand{\au}{\,\mathrm{AU}}

\newcommand{\kyr}{\,\mathrm{kyr}}
\newcommand{\yr}{\,\mathrm{yr}}

\DeclareRobustCommand{\VAN}[3]{#2}
\let\VANthebibliography\thebibliography
\def\thebibliography{\DeclareRobustCommand{\VAN}[3]{##3}\VANthebibliography}

\usepackage{graphicx}	
\usepackage{amsmath}	

\defcitealias{Vito}{Paper~I}

\title[Simulated Analogues II]{Simulated analogues II: a new methodology for non-parametric matching of models to observations}

\author[R. Al-Belmpeisi et al.]{
Rami Al-Belmpeisi,$^{1}$
Vito Tuhtan,$^{1}$
Mikkel Bregning Christensen,$^{1}$
Rajika Kuruwita,$^{1,2}$
and Troels Haugb{\o}lle$^{1}$\thanks{E-mail: haugboel@nbi.ku.dk}
\\
$^{1}$Niels Bohr Institute,
University of Copenhagen, {\O}ster Voldgade 5, DK-1350 Copenhagen, Denmark\\
$^{2}$Heidelberg Institute for Theoretical Studies, Schlo{\ss}-Wolfsbrunnenweg 35, 69118 Heidelberg, Germany\\
}

\date{Accepted XXX. Received YYY; in original form ZZZ}

\pubyear{2023}

\begin{document}
\label{firstpage}
\pagerange{\pageref{firstpage}--\pageref{lastpage}}
\maketitle

\begin{abstract}
Star formation is a multi-scale problem, and only global simulations that account for the connection from the molecular cloud scale gas flow to the accreting protostar can reflect the observed complexity of protostellar systems. Star-forming regions are characterised by supersonic turbulence and as a result, it is not possible to simultaneously design models that account for the larger environment and in detail reproduce observed stellar systems. Instead, the stellar inventories can be matched statistically, and best matches found that approximate specific observations.
Observationally, a combination of single-dish telescopes and interferometers are now able to resolve the nearest protostellar objects on all scales from the protostellar core to the inner $10\au$.
We present a new non-parametric methodology which uses high-resolution simulations and post-processing methods to match simulations and observations using deep learning. Our goal is to perform a down-selection from large data sets of synthetic images to a ranked list of best-matching candidates with respect to the observation. This is particularly useful for binary and multiple stellar systems that form in turbulent environments. The objective is to accelerate the rate at which we can do such comparisons, remove biases from hand-picking matches, and contribute to identifying the underlying physical processes that drive the creation and evolution of observed protostellar systems.
\end{abstract}


\section{Introduction}
With the advent of long wavelength interferometers like ALMA and NOEMA, observations with tens of AU resolution have become feasible for the nearest protostars. This gives detailed information about how stars are formed, and in recent years, observations with high sensitivity have revealed how streamers, outflow cavities, and connections to the larger scale filaments play an important role in regulating a non-steady mass flow to young stars from their surrounding gas reservoir \citep{murillo2022cold,garufi2022alma,valdivia2022prodige,alves_case_2020, ginski_disk_2021}. In order to reproduce the complex gas kinematic structures of observed filaments or large-scale inflows and outflows, only models employing global simulations that properly account for the connection from the molecular cloud (MC) scale to the accreting protostar can be used to understand the systems further \citep{Kuffmeier_2017,2016A&A...587A..59F,2023EPJP..138..272K}.

\begin{figure*}
    \vspace{-1cm}
    \centering
    \includegraphics[width=\linewidth]{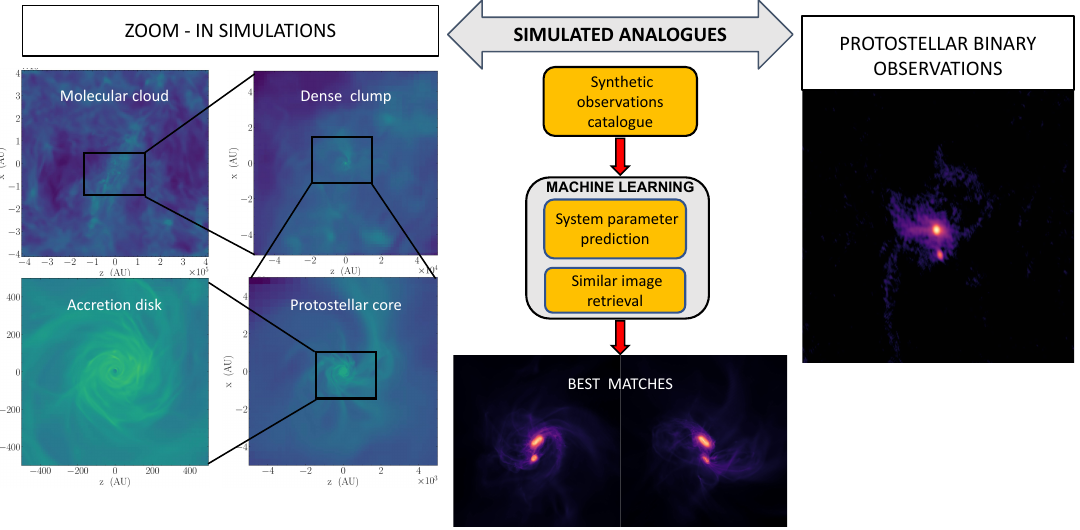}
    \caption{Schematic overview of the Simulated Analogues method for non-parametric matching of simulated systems to an observation. \emph{Zoom-in technique}: Target binary systems from molecular cloud simulations are selected for zoom-in simulations. \emph{Simulated analogues catalogue}: Synthetic images are produced from the zoom-in simulations, and system properties are derived (e.g. semi-major axis, masses, inclination etc). \emph{Machine learning}: The simulated analogues catalogue is partitioned into a test and training data set for the implemented DCNNs, and the best matches to observation are found.}
    \label{fig:fig1}
\end{figure*}

Recently, the problem of matching non-parametric simulations to observation has been addressed with a manual, labour-intensive procedure, which is not scalable to larger data sets  \citep{jorgensen2022binarity}. The interpretation of the observation was facilitated by finding simulation frames that reproduce similar structures to the observed line-of-sight moment-1 velocity maps. Synthetic moment-1 maps were constructed and outputs and observer viewpoints were identified that reproduced the observations of the binary system IRAS-2A. The analysis was instrumental in understanding the kinematics of IRAS-2A and framing the observation in a temporal context. It also highlighted the complex history of protobinary systems. 

Computing advances have allowed the era of machine learning with a significant impact on data-driven discovery in astrophysics. Deep Convolutional Neural Networks (DCNNs) have found several applications in analysing rich astrophysical data sets. Examples include inferring Warm Dark Matter particle masses from image field data obtained from N-body cosmological simulations \citep{rose2023inferring}, determining cosmological parameters from 3D light-cone data from dark matter halos \citep{hwang2023universe}, morphologically classifying radio galaxies \citep{brand2023feature}, and detecting pulsars from numerous candidate sources in the X-ray \citep{wang2021method} and radio wave regimes \citep{guo2017pulsar}. In the field of star formation, machine learning has been utilised to identify and characterise protostellar outflows \citep{Offner1, Offner2} and signatures of stellar feedback bubbles in molecular line spectra \citep{Offner3} observed from CO emission in nearby star forming regions.

In this paper we introduce the Simulated Analogues toolbox, with which we aim to automate the process of matching simulations with observations. We use pipelined data processing including creation of synthetic observations, and employ DCNNs for down-selection. We use the analysis and synthetic observations tools and pipelines introduced in \cite{Vito} (Hereafter referred to as \citetalias{Vito}). In \Cref{sec:methods}, we provide an overview of the Simulated Analogues Toolbox. We briefly describe the MHD simulations and production of synthetic images in \Cref{ssec:simulations} and \Cref{ssec:post_processing} respectively. In \Cref{ssec:catalogue} our data, the simulated analogues catalogue, are discussed, while in \Cref{ssec:deep_learning}, we explain in detail how Deep Learning is applied to finding matches between simulation as observation. In \Cref{sec:results} we present the results obtained from applying Deep Learning models to observations of IRAS-2A.  Specifically, in \Cref{ssec:pretrained_dcnn_mol} and \ref{ssec:pretrained_dcnn_cont}, we apply pre-trained models to image data from molecular and dust-continuum emission respectively. Later in \Cref{ssec:trained_results}, we train an ensemble of neural networks and employ them once again for the analysis of the dust-continuum emission. Finally, in \Cref{sec:discussion} and \ref{sec:conclusions}, we assess the effectiveness of the method, and its application on future data.

\section{Methods and data}
\label{sec:methods}

In our pipeline, zoom-in simulations \citep{Kuffmeier_2017} are used to create synthetic observations of young stars. In \Cref{fig:fig1} a flowchart for the method is shown, summarising the steps. The first step involves running zoom-in simulations on target binary star systems. The second step concerns producing synthetic observations and deriving physical properties to build the `Simulated Analogues catalogue'. This catalogue then serves for training and testing machine learning algorithms to match the simulations with the observations. The first two steps, the zoom-in model and building the catalogue, are described in \citetalias{Vito}. Here we will provide a brief overview in \Cref{ssec:simulations} and \ref{ssec:post_processing}, while the catalogue itself is described in \Cref{ssec:catalogue}. The remainder of this paper is dedicated to demonstrating the validity of using Deep Learning to match simulation and observation non-parametrically.

\subsection{Simulations - tracking star formation inside a molecular cloud}
\label{ssec:simulations}

We used `zoom-in' MHD simulations \citep{nordlundzoomin,Kuffmeier_2017} of three isolated binary stars, selected from a simulation of star formation and early evolution inside a turbulent molecular cloud with a realistic initial mass function \citep{haugbolle_stellar_2018} and multiplicity distribution \citep{KURUWITA2023}. The simulations were run with the adaptive mesh refinement, ideal MHD code \texttt{RAMSES} \citep{RAMSES2002, RAMSESUPGRADE}.

In the zoom-in models presented here, radiation and thermodynamics during the collapse of the protostellar core are approximated using a polytropic equation of state with varying gamma \citep{masunaga_radiation_2000}. We allow for the finest resolution of either $0.8\,\au$ or $3.1\,\au$ in a region which is $20\,000\au$ across. The minimum cell size outside this region is rapidly increased to 201 $\au$ and then gradually increased to 403, 806, 1611, 3223, and 6446 $\au$ outside distances of 0.0625 pc (13.000 $\au$), 0.125, 0.25, 0.5, and 1 pc respectively to create an onion-like resolution structure that allows the gas and magnetic fields to be adequately resolved when approaching the protostellar system. We store outputs from the evolution for up to $260\kyr$  with a cadence of $100\yr$. This way we can accurately track the early evolution of protostellar systems from a realistic starting condition. These models are also described in \citet{jorgensen2022binarity}. For a more in-depth description of the simulations additional details can be found in \citetalias{Vito}.

\subsection{Post-processing and extracting observables}
\label{ssec:post_processing}

\begin{figure}
	\includegraphics[width=\columnwidth]{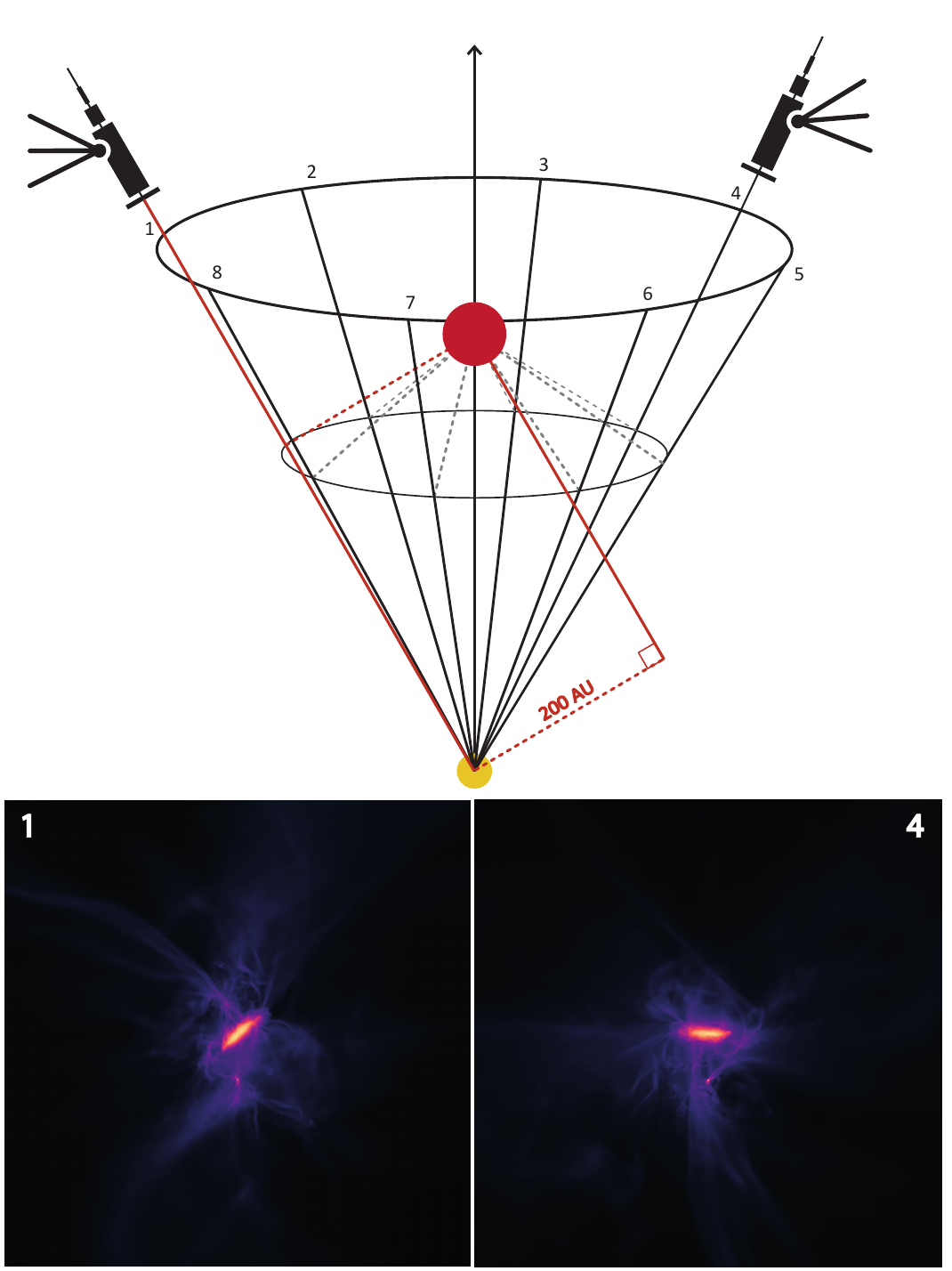}
\caption{Schematic of how lines of sight are derived from a binary, such that the projected separation equals $200\au$. Eight lines of sight are sampled evenly around the cone which gives the desired projected separation. The synthetic images below illustrate how the system would appear to observers located along the lines of sight 1 and 4 around the cone. The primary and secondary stars are indicated in red and yellow colours, respectively. Image Credit: Evangelia Skoteinioti Stafyla.}\label{fig:cone}
\end{figure}

To compare with observations, we construct synthetic images, spectral energy distributions and bolometric temperatures calculated for each simulation frame using RADMC-3D \citep{dullemond_radmc-3d:_2012} and convolved with a Gaussian to make the resolution comparable to the observations. This could in principle be done with any radiative transfer tool available in the community, and the effects of instrumentation could also be treated in more detail with e.g.~CASA \citep{casa}. For each snapshot of the simulations, we place the observer around a cone chosen such that the projected separation matches the separation of the observed protobinary, addressing the degeneracy that arises from projecting two points in 3D onto a 2D plane (see ~\Cref{fig:cone}). To provide a consistent frame of reference for interpreting the images, we set the north vector of the image plane such that it places the primary star exactly above the secondary star, just as in the observations. We sample the full 2$\pi$ angle by evenly distributing eight observers around the cone.

The simulated analogues catalogue consisting of synthetic images is expanded with physical quantities that describe the system, such as the protostellar masses, accretion rates, luminosities, and disk sizes. Details on extracting observables from the simulation can be found in \citetalias{Vito}. This catalogue is the data set used for both training and testing the inference using DCNNs.

\subsection{Synthetic observations catalogue}
\label{ssec:catalogue}
The simulated analogues catalogue used in this paper is generated by analysing all output from three simulated binary protostellar systems, namely B1, B2, and B3. The parameters of the simulated protostars can be found in \Cref{tab:zoom_in_sum}. Given an output cadence of 100 years, and eight observers, the resulting simulated analogues catalogue encompasses a total of $45\,000$ synthetic images for each observable. These include the dust-continuum emission discussed in \Cref{ssec:pretrained_dcnn_cont} and nine kinematic molecular tracers discussed in \Cref{ssec:pretrained_dcnn_mol}, giving in total $450,000$ images for the full catalogue. For each output, we attach the following metadata: stellar ages ($t_1$, $t_2$), masses ($M_1$, $M_2$), accretion rates ($\dot M_1$, $\dot M_2$), luminosities ($L_1$, $L_2$), disk sizes ($d_{size}$, $d_{size_2}$), the direction of the angular momentum of the circumstellar disks $\mathbf{J}_1, \mathbf{J}_2$ and disk masses ($M_{d1}$, $M_{d2}$) for the primary and secondary stars, respectively. We also include radial profiles of the gas computed as spherically binned mass weighted averages centred on each star for the angular momentum ($\vec{L}_1(r), \vec{L}_2(r)$) and radial velocity $\vec{v}_1(r), \vec{v}_2(r)$. Finally, we attach to each output the gas mass within a 100 AU sphere from the primary star, $M_{gas, r\leq100\au}$.

\begin{table}
\centering
\begin{tabular}{lccccr}
\hline
Name & $\Delta x_{min}$ & Final Age & M$_\textrm{prim}$ &  M$_\textrm{sec}$ & Periods \\
Units  & AU   & kyr      &  $M_\odot$        &  $M_\odot$       &    \\
\hline
\textbf{B1}  & 3.1       & 148          & 1.07 & 0.34      & 131      \\
\textbf{B2}  & 3.1       & 260          & 2.93 & 1.47      & 103      \\
\textbf{B3}  & 3.1       & 148          & 0.77 & 0.68      & 265      \\ 
\hline
\end{tabular}
\caption{ A short summary of the zoom-in simulations used in this work. $\Delta x_{min}$ is the smallest cell size on the highest level of refinement. The final age refers to that of the oldest star (usually the primary). The masses are measured at the end of the simulation. Periods correspond to the number of binary orbits a given system made measured from the first periastron crossing until the end of the simulation.}
\label{tab:zoom_in_sum}
\end{table}

\subsection{Deep Learning}
\label{ssec:deep_learning}

Deep Convolutional Neural Networks (DCNNs) are incredibly versatile mathematical models that have revolutionised the field of computer vision by significantly improving image recognition and analysis tasks. They are designed to mimic the visual processing capabilities of the human brain, allowing them to learn multiple levels of representation from data automatically. The hierarchical structure of DCNNs, with multiple layers of interconnected neurons, enables them to create progressively more abstract representations of the image data.

When a query image is fed into a DCNN, the output of the previous layer serves as the input for the subsequent layer, repeatedly passing it through a combination of mathematical operations such as 2D convolution along the image's spatial extent. Every layer is followed by an activation function which ensures non-linear mapping and generates the layer output. This representation is commonly referred to as the image's latent representation, or feature vector, or feature map. In the initial layers of a DCNN, the layer outputs typically consist of small-scale features which are characterised by high spatial dimensions. As the layers go deeper into the network, the features become progressively less spatially refined but can identify larger and more complex patterns. The feature maps from the final layers of the network are capable of abstracting high-level features and shapes, ultimately such as spirals or high contrast jumps in density. While training DCNNs requires large amounts of labelled data and computationally intensive processes, their impressive performance justifies the investment.

Using the last layer of a neural network configuration, we generate feature maps for all the images and compare them to the observation feature map using element-wise distance measures to assess a quantifiable similarity measure. This approach allows us to rank the entire list of synthetic images from least to most similar with respect to the observation. We can then select the most similar images to the observation by down-selecting to a list of best candidates, which serve as a sample for exploring patterns in the simulated observable and physical quantities and as a basis to put the observation into a dynamical evolutionary context.

Apart from the similar image retrieval algorithm, we also utilise feature maps as input to train a Multi-Layer Perceptron (MLP) on top of the pre-trained DCNNs. MLP is a feed-forward type of artificial neural network where the data flows from the input to the output layer in one direction through one or more "hidden layers" which are fully connected \citep{rosenblatt1958perceptron}. 
By combining a variant of the triplet loss scheme \Citep{schroff2015facenet} and a scoring function, we expand the capability of the neural network. The goal of training the MLP neural network is to connect prominent visual features with the binary systemic parameters to produce a model that retrieves the most similar images in terms of systemic parameters by design. 

In both approaches, the feature maps are compared to each other through element-wise distance measures. The aim is to retrieve simulation frames that share maximal visual similarity with the observation. We have explored the following distance metrics:
\begin{enumerate}
\item Cosine Distance
\begin{equation}
\label{eqn:cosinedistance}
    d_1(\vec{u},\vec{v})=1- \frac{\vec{u} \cdot \vec{v}}{|\vec{u}| |\vec{v}|}\,,
\end{equation}

\item Euclidean Distance
\begin{equation}
\label{eqn:Euclideandistance}
    d_2(\vec{u},\vec{v})=\sum_i (|u_i-v_i|^2)^{1/2}\,,
\end{equation}

\item Braycurtis Distance
\begin{equation}
\label{eqn:braycurtis}
d_3(\vec{u},\vec{v})=\sum_i\left|u_i-v_i\right| / \sum_i\left|u_i+v_i\right|\,,
\end{equation}

\item Canberra Distance
\begin{equation}
\label{eqn:canbera}
d_4(\vec{u}, \vec{v})=\sum_i \frac{\left|u_i-v_i\right|}{\left|u_i\right|+\left|v_i\right|}\,,
\end{equation}

\item L1  Distance \footnote{ L1 distance is also referred to as city-block distance or Manhattan distance.}
\begin{equation}
\label{eqn:cityblock}
d_5(\vec{u},\vec{v})=\sum_i\left|u_i-v_i\right| \,,
\end{equation}

\end{enumerate}

\noindent where $\vec{u}$ and $\vec{v}$ are the latent representations of the images that are being compared.

In order to systematically match an observation to the members of the simulated analogues catalogue, several DCNN were implemented using Python supplied with the packages of Tensorflow \citep{tensorflow} and Keras \citep{keras}. 
 
\subsubsection{Similar image retrieval using pre-trained convolutional layers}
\label{ssec:sir}

In the initially implemented approach, the used neural networks were initialised by loading the weights from the training on the ImageNet dataset \citep{deng2009imagenet}. The used networks include DenseNet121 \citep{huang2017densely}, EfficientNetB7 \citep{tan2019efficientnet}, InceptionV3  \citep{szegedy2016rethinking}, MobileNet \citep{Howard_2019_ICCV}, ResNetRS200 \citep{he2016identity}, Xception \citep{chollet2017xception}, VGG16 and VGG19  \citep{simonyan2014very}. Despite using pre-trained features on a different application domain, the DCNNs are expected to be a powerful tool in retrieving a statistical sample of the most similar images which can later prove useful in characterising several aspects of the observed systems.

This method employs a DCNN to extract feature maps from both the simulated images and the query image. These feature maps are then compared using the distance metrics mentioned in Equations \ref{eqn:cosinedistance}--\ref{eqn:cityblock} to determine the visual similarity between the two images. The similarity ranking is obtained by sorting the feature maps according to their computed distance measures, smaller distances for similar images and vice versa. This approach allows for the efficient identification of the simulated images that are most similar to the query image.

\subsubsection{Transfer Learning}
\label{ssec:transfer}

For the results of this section we used the following DCNNs:   EfficientNetB7 \citep{tan2019efficientnet}, MobileNet \citep{Howard_2019_ICCV}, ResNetRS200 \citep{he2016identity}, VGG16 and VGG19  \citep{simonyan2014very}. In this approach, we do not use the pre-trained neural network models as-is, but add additional trained layers on top them.

To effectively train a neural network it is essential to tune its hyperparameters, which refer to model parameters that are not learned directly from the data, but inserted manually by the user before the training. We use a common technique to optimise our model's hyperparameters, grid-search, which results in a model with the Adam optimiser \Citep{kingma2014adam}, which is configured with: $lr=10^{-4}, b_1=10^{-3}, b_2=0.999, \varepsilon=10^{-7}$. In order to create neural networks specialised for astrophysical images, we expand the neural networks by incorporating the MLP network. Aiming for simplicity, the implemented MLP consists of the input layer, only one hidden layer and the output layer with sizes of 256, 128 and 100 neurons respectively. 
 The MLP takes synthetic image data as input, derived from the output of the used DCNN, along with metadata which contain the physical parameters of the system. The objective of this implementation is to further specialise the network from \Cref{ssec:sir} in connection to the system's physical parameters in an astrophysical context. To prevent overfitting, while using large neural networks, we employ transfer learning techniques. This involves "freezing" the weights of the DCNN  rendering them non-updatable during back propagation, thus effectively training exclusively the MLP network. However, the novelty of this method relies on the interplay of two components: the \emph{triplet loss} and the \emph{scoring function}.

 The triplet loss scheme is an effective training method in deep learning, where a triplet of synthetic images (1, 2, 3) with known metadata parameters are randomly selected at each step of the training process. By default, the first member of the triplet is chosen to be the anchor ($a=1$). The two remaining analogues (2, 3) are compared with respect to the anchor based on the triplet metadata parameters. To achieve this, a scoring function, $g(q_i,q_j)$, is defined to measure the similarity between the metadata parameters $\vec{q}=({M_1, M_2, L_1, L_2, d_{size}, d_{size_2}, M_{gas, r\leq100\au}})$ of a pair of simulated analogues ($i,j$). The scoring function is constructed to satisfy the specific constraints: 
 \begin{enumerate}
     \item if two images are similar in terms of metadata parameters, the scoring function should be small, close to zero;
     \item if two images differ significantly in terms of metadata parameters, the scoring function should be large;
     \item The scoring function should be non-negative.
 \end{enumerate} 
 
The simulated analogue with the lowest scoring function is labelled as positive (p), while the analogue with the highest scoring function is labelled as negative (n). The suggested scoring function is calculated element-wise for each component of metadata $q$ by dividing the absolute difference of the metadata parameters by the sum of the metadata parameters:

\begin{equation}
    g(q_a,q_j)=\frac{|q_a-q_j|}{q_a+q_j}\,,
\end{equation}
where the index $j$ takes values from the two remaining members of the triplet set of images (2,3).

A score which takes into consideration the scoring functions is then used to determine which image is considered positive or negative in the training process. The score is defined as the difference in scoring functions of the two competing members of the triplet (2,3) with respect to the anchor (a):

\begin{equation}
     Score = \sum_{i=1}^{k}(g(q_a^i,q_2^i)-g(q_a^i,q_3^i))\,,
\end{equation}
where the summation over $i$ is performed over all the metadata parameters to get the final score. As earlier mentioned, since our metadata consists of 6 systemic parameters for the description of a binary system, $k=6$. Given a triplet of synthetic images along with their metadata, the score calculation proceeds, which sets the positive (p) and negative (n) analogue according to the following relation:

\[ 
(p,n)= \left\{
\begin{array}{ll}
      (2,3) & Score > 0 \\
      (3,2) & Score < 0 \\
      \text{Undefined, resampling triplet} & Score=0
\end{array} 
\right. 
\]

The mathematical expression for the triplet loss algorithm is usually expressed in terms of the following quantity:
\begin{equation}
    \left[\left\|\vec{f^{a}}-\vec{f^{p}}\right\|-\left\|\vec{f^{a}}-\vec{f^{n}}\right\|\right]_{+}=\left[\left\|d_{ap}-d_{an}\right\|\right]_{+}\,,
\end{equation}
where $\vec{f^{a}}$,$\vec{f^{p}}$ and $\vec{f^{n}}$ are the 100-dimensional feature maps from the output of the MLP for the anchor, positive, and negative image respectively and $d_{ap}$, and $d_{an}$ are the $L_1$ distances between the anchor image and positive and negative scoring images respectively. If the $L_1$ distance between the anchor and positive $\left\|\vec{f^{a}}-\vec{f^{p}}\right\|$ is smaller than the $L_1 $ distance between the anchor and negative $\left\|\vec{f^{a}}-\vec{f^{n}}\right\|$, then the difference of those terms will be negative and the total result will be zero; no learning will happen at this training step, since for this triplet the neural network is already predicting 100-D feature maps which are closer for the pair (a-p) compared to the pair (a-n). However, if the opposite was true, and the negative feature map $L_1$ distance is closer to the anchor than the positive is, learning is applied in this training step according to the following algorithm: the two distances are transformed into a probability distribution with the Softmax function, which is defined as
\begin{equation}
P(p)=\frac{e^{d_{a p}}}{e^{d_{a p}}+e^{d_{a n}}}\,,
\end{equation}
\begin{equation}
P(n)=\frac{e^{d_{a n}}}{e^{d_{a p}}+e^{d_{a n}}}\,.
\end{equation}
After the probabilities are computed, the probability $P(n)$ of having a larger distance with respect to $a$ than $p$ does is maximised (i.e. $d_{a n} \gg d_{a p}$), through the minimisation of the loss function $Loss = P(p) +(1-P(n))$.

To accurately evaluate the model, we split the aforementioned code into eight projections and use the ones with $\varphi = (0, \pi/2, \pi, 3\pi/2)$ for training and the projections with $\varphi = (\pi/4, 3\pi/4 , 5\pi/4 ,7\pi/4)$ for testing. We employ regularisation strategies, which are used in machine learning to reduce overfitting and avoid our model performing poorly on unseen data. L2 regularisation includes a penalty term based on the L2 norm of the weights in the MLP layers. Large weights are discouraged by this penalty, resulting in more generalisable representations within the network. L2 regularisation can be mathematically defined as:

\begin{equation}
\text { Loss }^{'}=\text { Loss }+\lambda \sum_{i=1}^n w_i^2\,,
\end{equation}

where $w_i$ are the weights of the network and $\lambda = 10^{-5}$ is the tuning parameter that sets the amount that we want to penalise large weights within the MLP network. We use the initialisation of \citet{he2015delving} to set the initial values for the weights in the MLP layers, which allows deep networks to efficiently converge. As a last approach to address overfitting, we use dropout \citep{srivastava2014dropout},  a technique that involves assigning a probability, denoted as P, for the presence of a layer of neurons/neuron during a training step. We implement 2D spatial dropout when passing the feature maps from the DCNNs to the MLP. Additionally, dropout is applied to the first two layers of the MLP. For all approaches, we use $P=0.5$.

The objective of this transfer-learning design is to develop a machine learning model capable of assessing the similarity between observed protostellar systems based on their images and physical parameters. To accomplish this, we employed a variant of the triplet loss training algorithm that maps the DCNN-produced feature maps through the MLP to generate a new trained 100-D feature map. These 100-D latent representations are constructed to contain a similar encoding for query images of protostellar systems with similar physical parameters, through the non-linear mappings of a MLP network and the incorporation of metadata. The resulting model allows for the comparison of protostellar systems based on image and parameter similarity, facilitating further understanding through their statistical properties.

\begin{figure*}
\includegraphics[width=0.75\linewidth]{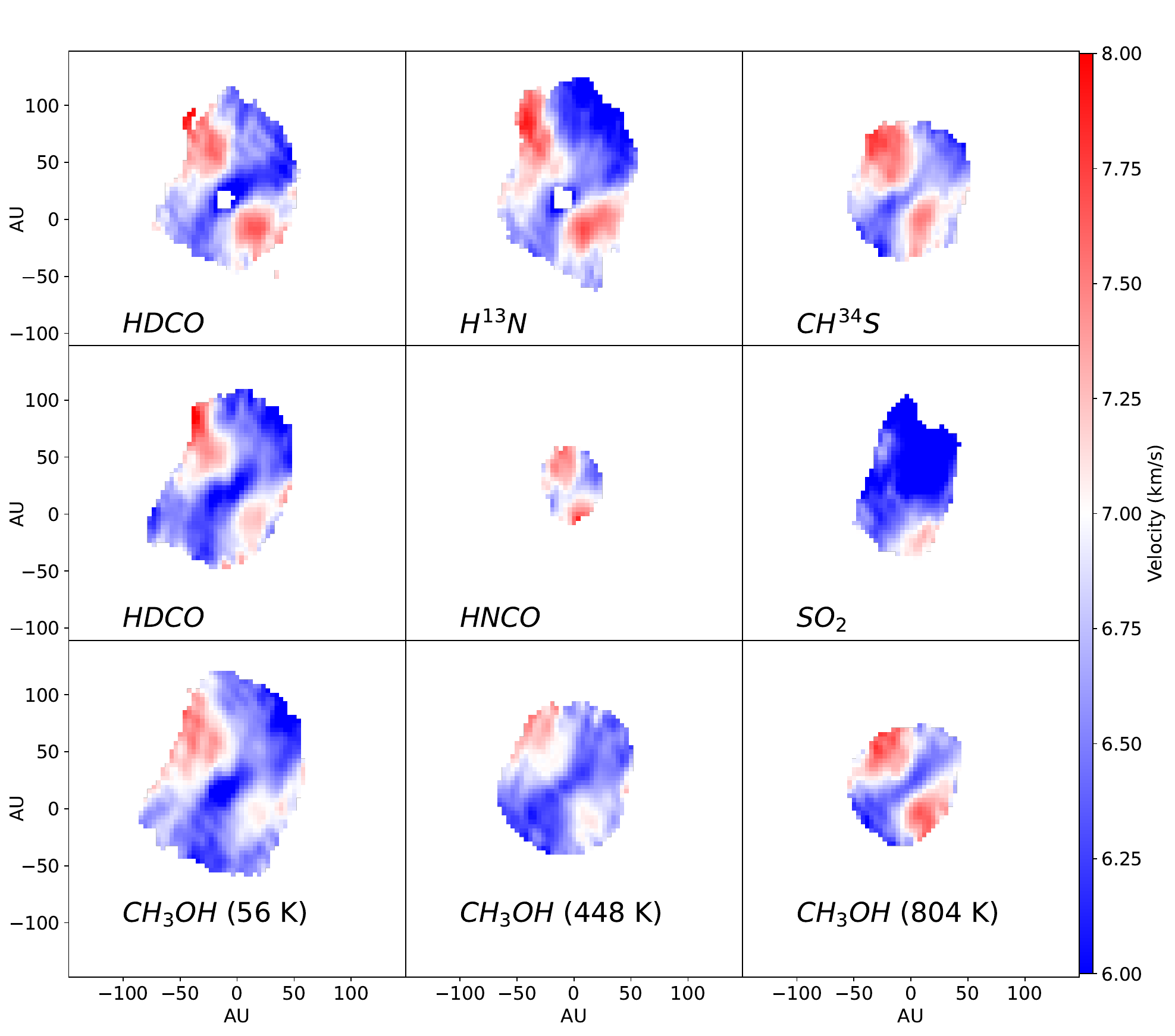}
    \caption{Moment-1 maps close to the primary star of IRAS-2A. In each panel, the transition lines are integrated across velocities between 2 and 12 $km$ $s^{-1}$ ($\pm 5 km$ $s^{-1}$ relative to the systemic velocity) at positions where the integrated emission exceeds 10 $\sigma$. The figure shows a selection of transition lines for complex organic molecules and displays consistently a quadrupole velocity structure. These data are taken from \citet{jorgensen2022binarity}. }
        \label{fig:moment_maps}
\end{figure*}

\section{Results}
\label{sec:results}
In this section, we use both generic pre-trained DCNNs and DCNNs specifically trained for our purpose to identify good matches between synthetic images in the catalogue and the high-resolution observation of the protostellar binary IRAS-2A. We use both line and continuum tracers at different scales to match both the small-scale kinematics in the immediate vicinity of the primary and the larger-scale environment mapped in the continuum. Ultimately, these matches can then be used to extract probable metadata parameters that inform us about the age, evolutionary stage, dynamics and physical parameters of the observed system.

\begin{figure*}
    \centering
    \includegraphics[width=0.8\linewidth]{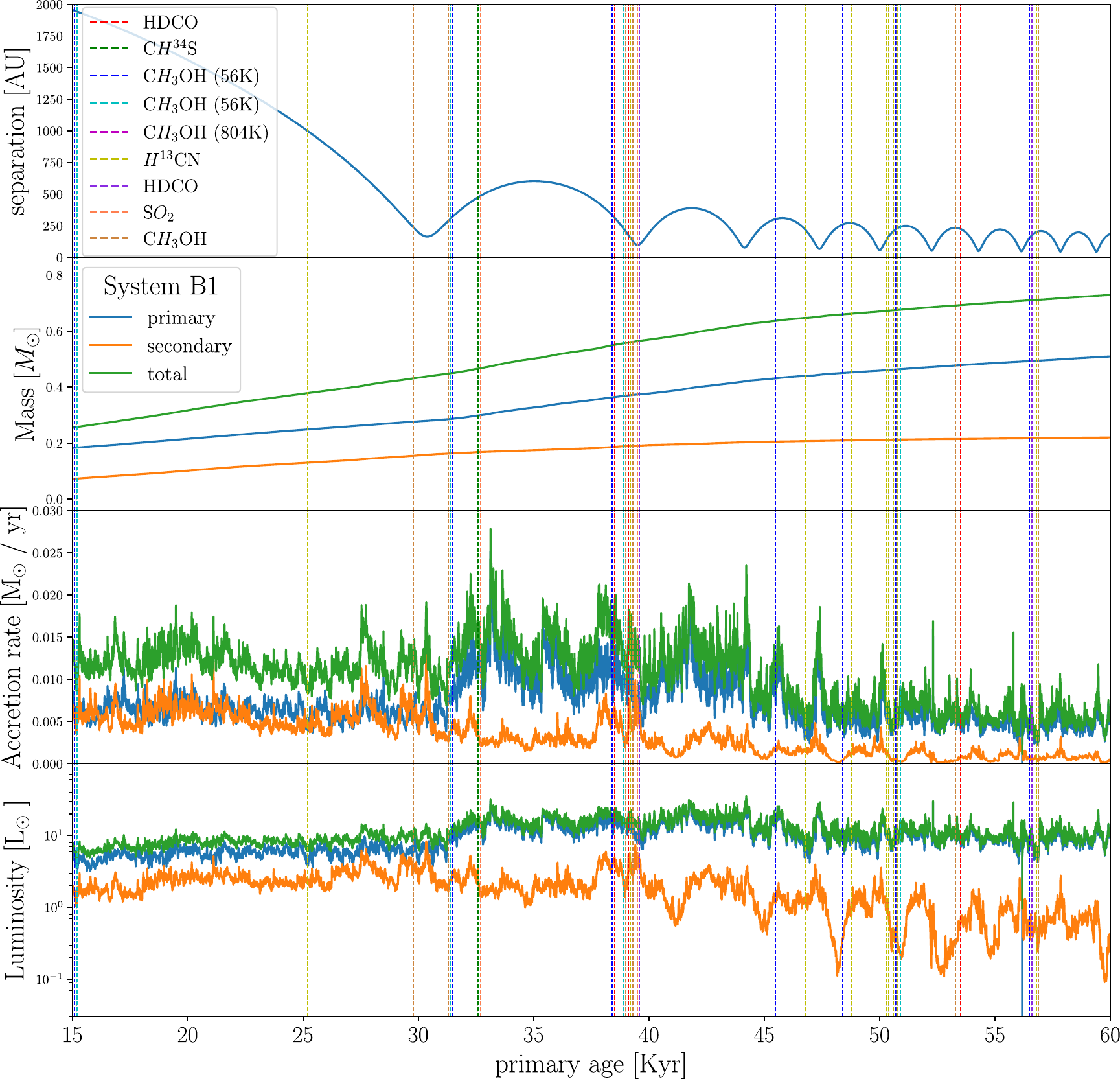}
    \caption{Parameter evolution plots for system B1 \citepalias{Vito} showing separation, mass, accretion rate, and luminosity. The vertical lines indicate the respective time that each best match was observed using VGG16 and VGG19 on the molecular line emission (\Cref{ssec:pretrained_dcnn_mol}), while the colour reflects the molecule moment map that produced the best match as can be seen in the legend.}
    \label{fig:bestmatches}
\end{figure*}

\begin{figure*}
    \centering
    \includegraphics[width=\linewidth]{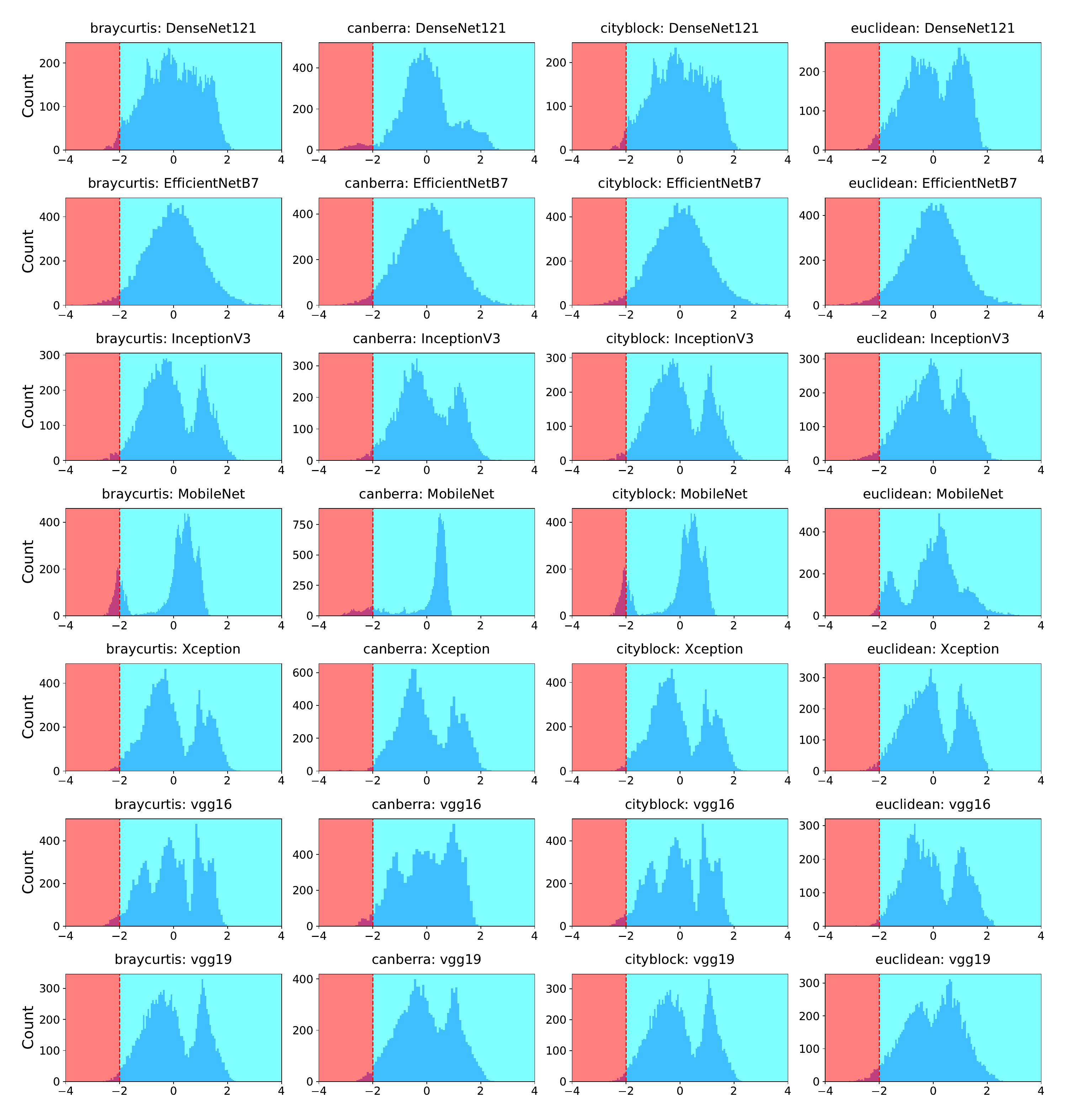}
    \caption{Distance distributions of each combination of pre-trained DCNNs and distance metrics in \Cref{ssec:pretrained_dcnn_cont}. The distances are scaled to unit variance ($\sigma$) and zero mean. The simulated analogues in the red regions represent matches with the smallest distances to the observations and have a distance that is at least $2\sigma$ from the mean distance.}
    \label{fig:histogram}
\end{figure*}

\subsection{Pre-Trained DCNNs for Image Retrieval: moment-1 maps}
\label{ssec:pretrained_dcnn_mol}
Moment-1 maps derived from molecular tracers are a powerful tool for understanding the kinematic structure of young protostars. Different tracers have different critical densities, optical depths, and reaction pathways allowing for mapping at different scales of the system. Complex kinematics and corresponding moment-1 maps resulting from a combination of (filamentary) inflow, outflows, and a nascent disk are observed and expected based on modelling. This is particularly true for binary and multiple systems. \citet{jorgensen2022binarity} observed the protostellar binary IRAS-2A and derived nine moment-one maps at the 100 au scale, based on transitions in different molecular tracers (see \Cref{fig:moment_maps}).

We use the pre-trained convolutional networks on this data set to explore how well the feature maps can be used to automatically extract similar images from incomplete data sets covering relatively few resolution elements in a small field of view. In general, using DCNNs on incomplete data is challenging and has only recently been addressed \citep{9706982}. However, as one can see from the observed moment-1 maps shown in \Cref{fig:moment_maps}, the emission is coming from specific regions. To overcome this issue, we first rotate, then zoom, and finally crop a square region from each moment map image separately such that only a few areas in the processed image have empty values. This could also be accomplished by including the missing data directly from the observations, if we have had access to the raw data. Afterwards, we perform 2D nearest-neighbour interpolation to approximate the missing pixels. As a last step, we use cubic interpolation to scale the image to a resolution of $224^2$ pixels, which matches the original input size of the neural networks. Note that the neural network can be scaled to analyse higher resolution images, with the cost of significant up-scaling of feature map sizes, and training time.

The cropping and rotation-specific parameters that are used for each moment-1 map are also applied to the simulation outputs, to produce synthetic moment-1 maps that capture the same scales (cropping), and orientation (rotating) as the moment maps with respect to the binary system in the observations.

In this initial application, for simplicity, we experiment with the VGG16 and VGG19 architectures, due to their relatively simple feed-forward architecture, in contrast to more recent ones, such as ResNet which incorporate residual connections between non-neighbouring layers. Residual connections are empirically known to effectively ease the optimisation procedure during back propagation, resulting in models that have additional shortcut identity connections without increasing model complexity or including extra parameters \citep{he2016deep}.

After experimenting with the different distance measures we have chosen to define the distance $D_{ij}$ between two feature vectors of the $j^{th}$ observed moment-1 map ($\vec f^{j}$) and the $i^{th}$ synthetic observation ($\vec f^i_{SO}$) is computed as the squared sum of normalised Cosine and Euclidean distances
\begin{equation}
    D_{ij}=d_1^2(\vec f^{j},\vec f^i_{SO})/N_{ij1} +d_2^2(\vec f^{j},\vec f^i_{SO}) / N_{ij2}\,,
\end{equation}
where $N_{ijk} = \max (d_k(\vec f^{j},\vec f^i_{SO}))$.

In \Cref{fig:bestmatches} we show the 50 synthetic images with the lowest distance to each of the nine moment-1 maps as vertical lines. We see clear clustering of the different best matches, and cross consistency, with different molecular line maps resulting in similar best matches. The best matches are similar though not identical to the best matches identified by hand by \citet{jorgensen2022binarity}, which serves as an independent validation of the approach.

\subsection{Pre-Trained DCNNs for Image Retrieval: dust-continuum emission}
\label{ssec:pretrained_dcnn_cont}
To evaluate the performance of DCNNs on large-scale data sets with many resolution elements we have applied the methodology to find simulated matches to the dust-continuum observation of the protostellar young stellar IRAS-2A on the thousand AU scale.
Additionally, we upscaled the networks to analyse images of $800^2$ resolution which results in a pixel size of $2.5 \au$.

\begin{figure*}
    \centering
    \includegraphics[width=0.76\linewidth]{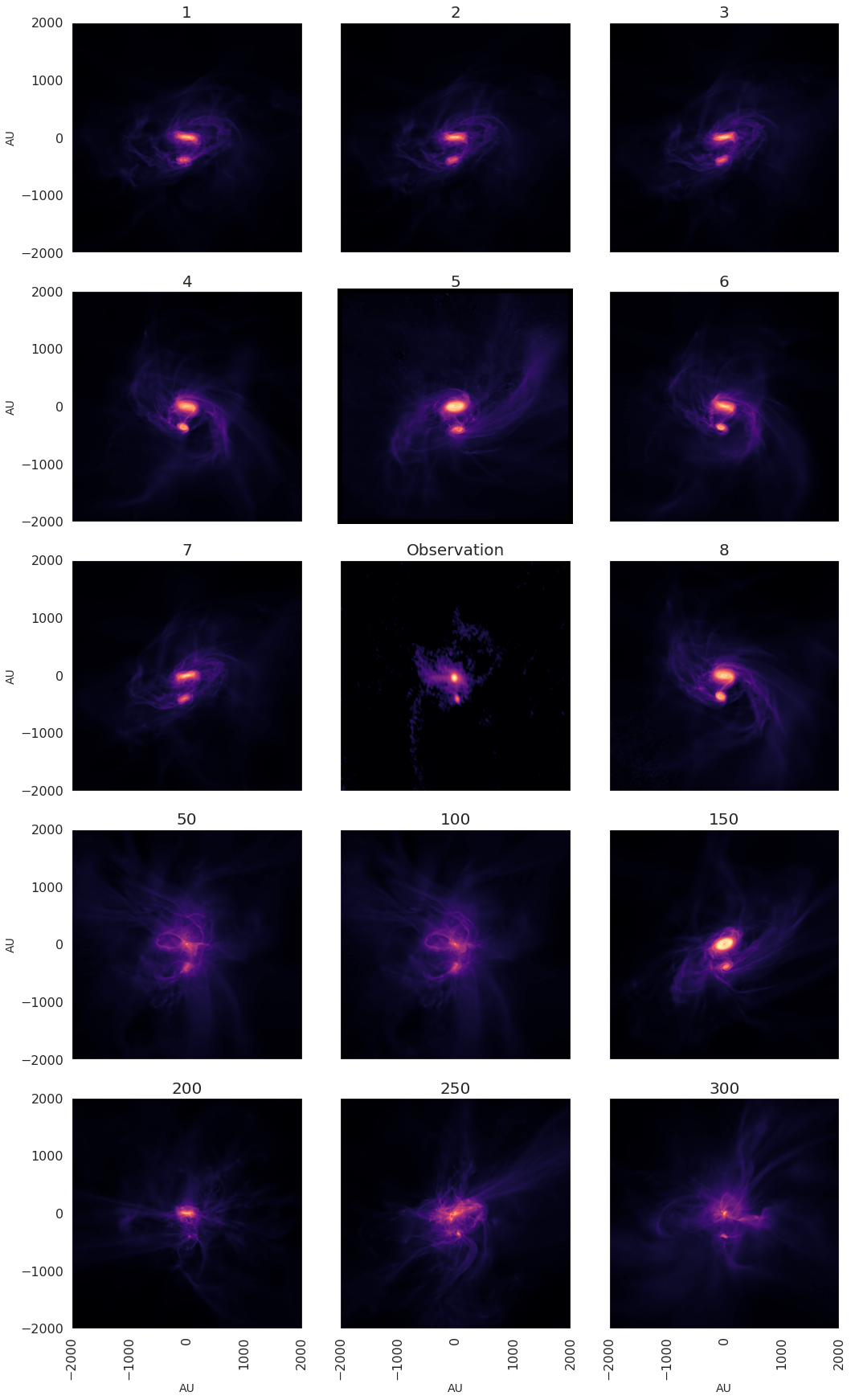}
    \caption{Ranked selection of synthetic observations and the continuum image of IRAS-2A \citep{jorgensen2022binarity}. IRAS-2A is shown in the middle panel of the third row. The ranking of the images is annotated above each panel. 
    The ranking is determined using DenseNet121 and an Euclidean distance metric.}
    \label{fig:9BESTMATCHES}
\end{figure*}

As for the moment-1 maps, each network ranks the synthetic images in terms of similarity by measuring the distance between the corresponding feature vector and the feature vector of the observed continuum flux. \Cref{fig:histogram} shows the histogram distribution of distances shifted to the average value and normalised by the standard deviation. A priori this abstract similarity measure for the feature vectors does not necessarily correspond to a similarity of the synthetic images to the observed image. To test this, we evaluated the similarity of the 8 best-ranked images for each combination of deep neural networks and distance metrics subjectively. From this, we conclude that the Euclidean distance measure applied to the feature vectors produced by DenseNet121 is the best combination for selecting images that are similar to the observation, as shown in \Cref{fig:9BESTMATCHES}.

\begin{figure*}
    \centering
    \includegraphics[width=1\linewidth]{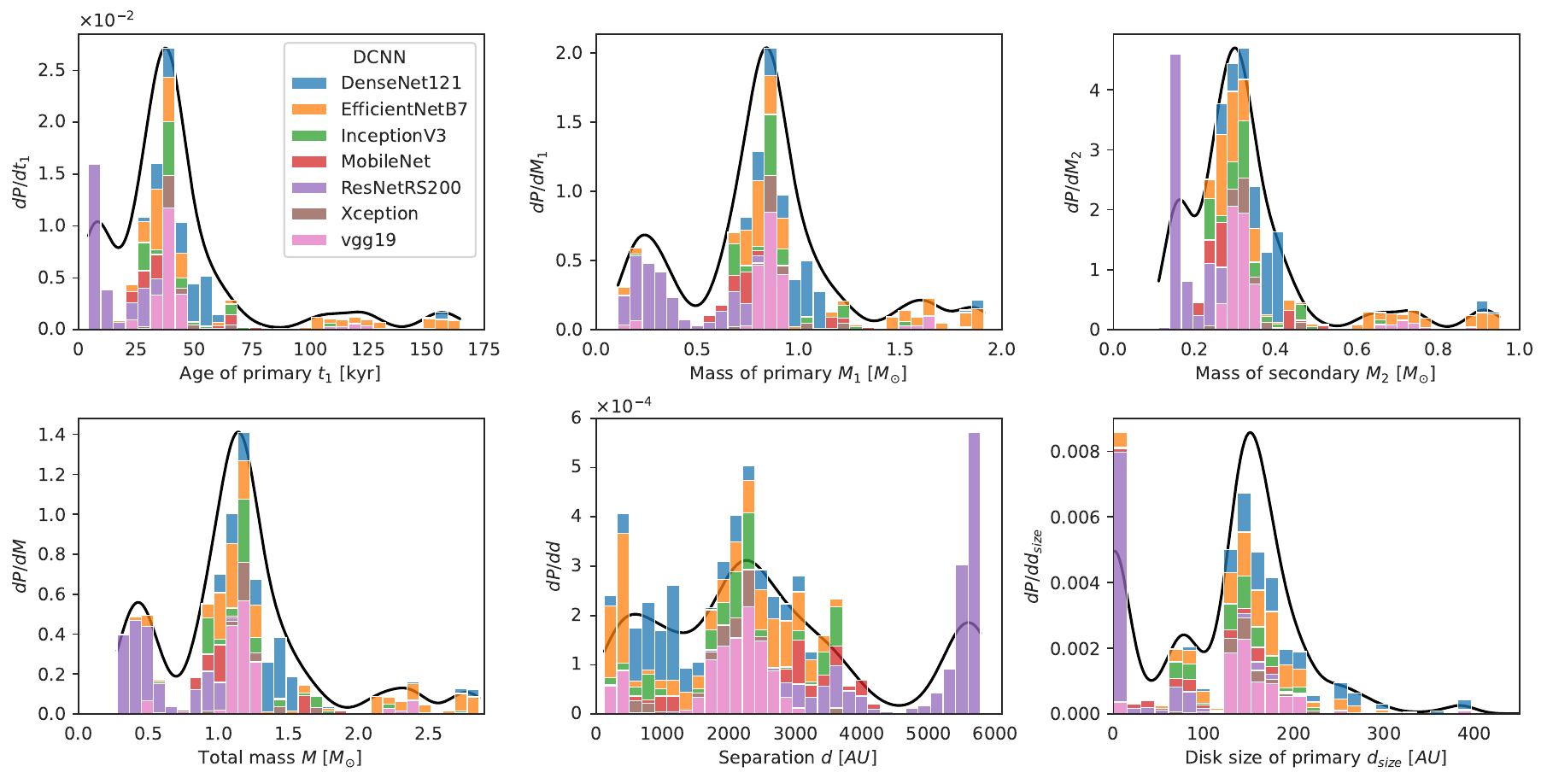}
    \caption{Stacked histograms and probability density functions for the predicted system parameters using pre-trained DCNNs as described in \Cref{ssec:sir}. The parameters are derived from the synthetic observations that are in the red region in  \Cref{fig:histogram}. We use the DCNNs listed in the legend together with an Euclidean distance metric. The overall variations of the histograms (black line) are captured with a kernel density estimator. The predicted parameters, comprising the age of the primary $t_1$, mass of the primary $M_1$, mass of the secondary $M_2$, total mass $M$, binary separation $d$, and disk size of the primary $d_{size}$, are arranged from top-left to bottom-right.}
    \label{fig:all_kde}
\end{figure*}

\begin{figure*}
    \centering
    \includegraphics[width=1\linewidth]{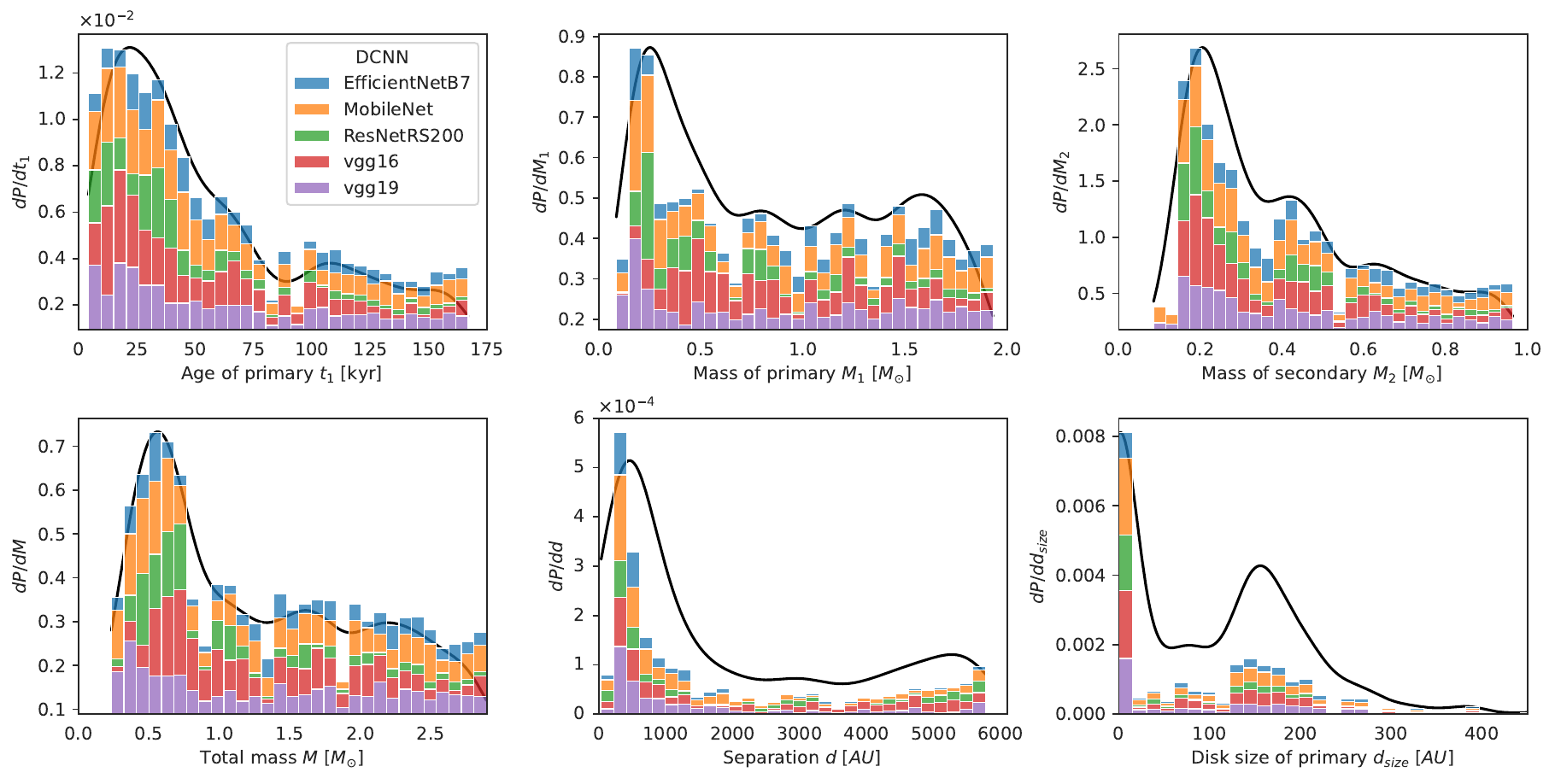}
    \caption{Stacked histograms and probability density functions for the predicted system parameters using the trained DCNNs which are augmented with the MLP layers described in \Cref{ssec:trained_results}. The parameters are derived from the synthetic observations that are at least $-2\sigma$ away from the mean distance (similar to the red region in \Cref{fig:histogram}). We use the DCNNs listed in the legend together with an Euclidean distance metric. The overall variations of the histograms (black line) are captured with a kernel density estimator. The predicted parameters, comprising the age of the primary $t_1$, mass of the primary $M_1$, mass of the secondary $M_2$, total mass $M$, binary separation $d$, and disk size of the primary $d_{size}$, are arranged from top-left to bottom-right.}
    \label{fig:kde_trained}
\end{figure*}

Based on the ranking, we create a system parameter estimator tool that uses the metadata of the best-ranked simulation outputs to construct a distribution of plausible derived parameters for the observation. We have made a cut by choosing all simulation outputs that have a distance in feature vectors from the observation that is less than -2$\sigma$ (see \Cref{fig:histogram}). The distribution of these parameters
are shown as histograms and Kernel Density Estimators (KDE) for the binary total mass ($M$), mass of the primary ($M_1$), mass of the secondary ($M_2$), age of the primary ($t_1$), separation ($d$), and disk size of the primary ($d_{size}$) respectively in \Cref{fig:all_kde}.

Apart from an outlier generated by ResNetRS200, we see that the results cluster around a well-defined peak corresponding to an age interval of 25 to 60 kyr. Typically, the true distance is similar to the observed distance though for some simulation outputs, there is a high angle between the image plane and the orbital plane. The disk size of the primary is predicted to be somewhere between $140\au$ and $200\au$ and the total mass is between 1.0 and $1.5\msun$.

\subsection{Trained DCNNs for Image retrieval: scoring and triplet loss}
\label{ssec:trained_results}

In the previous sections we have used DCNNs that are designed to distinguish between different images but optimised for astrophysical images. Here we introduce the technique of similar image retrieval, which includes the use of metadata scoring and triplet loss, to train a neural network. During the training process, each network completes 10 epochs, each of which involves loading 18000 triplets of images in batches of 64 triplets at a time for each weight update.  The training of each network takes approximately 5 hours on a Nvidia A100 GPU. The networks learn by minimising the loss function. 

Based on the neural network ensemble predictions, we implement the system parameter estimator tool that was previously mentioned in \Cref{ssec:pretrained_dcnn_cont} which produces probability density functions for the predicted system parameters, as shown in \Cref{fig:kde_trained}. These results consist of analogues within the $2\sigma$ best matches.

\section{Discussion}
\label{sec:discussion}
Systematic matching of observations to simulations is necessary to properly take into account degeneracies produced by projection effects and time evolution. The method aims to determine which members of the simulated analogues catalogue better describe a given observation. Essentially, the problem we are solving is finding the point in time and viewing angle from which our simulations best match the observation. ``Similarity to observations'' is difficult both to define and quantify.  Nevertheless, using our neural network algorithm, we can reduce the number of possible images from tens of thousands to a handful of simulated analogues based on either their visual similarity alone or combining their visual similarity and systemic parameters. Analogues such as those described in \Cref{ssec:pretrained_dcnn_mol} and \Cref{ssec:pretrained_dcnn_cont}, are ranked in terms of visual similarity, i.e., gas morphology. This can serve as a useful tool for interpreting recent observations, which reveal an increasingly complex variety of gas structures around YSO binaries \citep{zhang_dynamics_2019, kumar_unifying_2020, alves_case_2020, ginski_disk_2021}. Our neural network predictions for the age of the primary could also be used as a constraint on the evolutionary class of each observed system, according to the class-age relation that is described in \citet{dunham2015young}. Currently, a major bias and uncertainty is the underlying simulations catalogue. Using zoom-in models is important for creating a catalogue with a realistic core structure, infall history, and anchoring of magnetic fields; this is particularly the case for binary or multiple systems. However, currently, the available data set is limited by the number of such zoom-in models. In this paper, we have chosen to use only three models to compare with IRAS-2A. This could be seen as a major bias, but fortunately, these systems were already chosen to match the overall system parameters of the observation \citep{jorgensen2022binarity} and therefore well-suited for such a comparison.

To ensure the accuracy and reliability of deep learning models within the Simulated Analogues toolbox, it is crucial to have access to large and diverse data sets that capture the evolution of numerous simulated systems originating from different regions of the molecular cloud. This diversity is necessary to capture the full range of conditions under which single, binary, and multiple young stellar objects (YSOs) form and to enable the models to derive more precise conclusions on the system parameters. The focus in this paper is on binary YSOs, where the classical after-the-fact modelling of the collapse of an isolated core is challenging at best, and the methodology is particularly well-suited. But the method is general. We anticipate that Deep Learning models within Simulated Analogues can be successfully extended and applied to assess the similarity of simulations and astrophysical observations of all kinds, as long as appropriate databases are available that contain both image data and systemic parameters.

The IRAS-2A system observation contains some striking visual features, those of the spiral trails which are located at the southwest and northeast. Only some of the deep learning configurations retrieve images which also appear similar to our eyes, while most of the combinations of distance measures and DCNNs produce matches that do not appear to be as useful for the task of image retrieval. Several neural networks like the pre-trained Densenet121 can down-select to a handful of synthetic images which are similar to the observation both at the centre and in the extended structure, like the bulge-like feature west of the binary system, the connecting bridge between the young stellar objects or the spiral trailing gas structures. Additionally what is interesting to observe is that DenseNet121 can capture structures that share similar spiral trailing features but counter-rotating. This validates that it is a network which displays a high degree of translational, rotational, and mirroring invariance in its features. 

Focusing on small-scale features and using Deep Learning matching with VGG16 and VGG19 for the molecular line emission (see \Cref{ssec:pretrained_dcnn_mol}), the models predict matches which cluster around distinct short periods throughout the evolution, mainly around 38, 51 and 56 kyr, as shown by the vertical dashed lines in \Cref{fig:bestmatches}. However, most of those best matches are found in clustered regions within the $30-60\kyr$ interval. 

The results of \Cref{ssec:pretrained_dcnn_cont} demonstrated in \Cref{fig:all_kde} show more variety. The distribution of prediction system parameters, such as the age of the primary, are similar for most DCNNs, but there are some outliers. As seen in \Cref{fig:all_kde} ResNetRS200 picks up images corresponding to a very early evolutionary state when the two stars are still approaching each other, while EfficientNetB7 picks up best matches at very late times. The majority of the networks cluster around the ages of 25 to 60 kyr, regardless of the exact DCNN used. The shape of the PDF with respect to the age of the primary is reflected on the PDFs of $M_1$, $M_2$ and $M$ since mass is increasing monotonically with time. On the other hand, the separation between the protostars ($d$) and the disk size ($d_{size}$) follow different temporal dependencies which suggest that the most probable 3D separation between the two stars is around $\approx2300\au$ and the most probable circumstellar disk size for the primary is estimated to be $100-200\au$. 

Lastly, using a trained network, where the metadata similarity is taken into account, \Cref{fig:kde_trained} shows that the system IRAS-2A is most likely less than $75\kyr$ old. The individual DCNNs produce similar results, and the best matches have a broader distribution in all system parameters. One notable difference compare to the untrained results is that the network preferentially find matches where the stars are closer to periastron or times where the system has already dispensed their orbital angular momentum to the gas and have separations smaller than $1000\au$. While we still observe a local maximum on the probability density distribution of the circumstellar disk size of the primary at $150-200\au$, there is also a strong peak at epochs where no disk is present ($d_{size}=0$).

In conclusion, according to our current small database, most methods qualitatively agree and suggest that the age of the primary of IRAS-2A  $\sim30-60\kyr$. However, there is variation in the predicted system masses and separations, which depends on the neural network configuration and training methods used. The obtained results are in good agreement with a more observational approach and taking into account other observational indicators, such as the class 0 classification of the system, the absence of disk detection and the strong and misaligned outflows. 

\section{Conclusions}
\label{sec:conclusions}

In this paper we have introduced a novel method to efficiently process massive simulation data sets with the end goal of identifying good matches with observed YSOs.

As a first pilot application, we have used machine learning and neural networks to characterise the age and system properties of the YSO binary IRAS-2A using simulations of binary star formation. Based on our analysis, this YSO binary can be characterised as a young protostellar system, which is consistent with \citet{jorgensen2022binarity}. The predictions from our Deep learning models, qualitatively agree that the system IRAS-2A is most likely less than $60\kyr$.

This is a first pilot study, and the predictions for the stellar parameters should not be over-interpreted, given that the database of simulated analogues on which the models were trained is limited, and we are still exploring the best practises for how to use machine learning to identify best matches. The evolution of YSO system parameters can be significantly influenced by the kinematic structures and gas flow present in the interstellar environment where the simulated system is located. This suggests that by including additional simulated binary systems, the metadata parameter space can be expanded, as more evolutionary curves of systems are added. Expanding to tens or hundreds of simulated systems is crucial in building a robust and versatile database that enables machine learning models to achieve generalisation rather than over-fitting the limited available data, and for automating the process in conjunction with e.g.~larger observational surveys such as eDisk \citep{eDisk}. While constructing models using a data set that consists of only three systems and their corresponding evolution, we expect that these models will perform well only in analysing observations that correspond to similar physical conditions to our simulations. 

In this paper, we have shown the viability and potential of the method, but much still has to be done in terms of characterising the different DCNNs and their applicability to astronomical images. 

We expect that in the coming years, several generic DCNNs will be developed, which will enable machine learning models to extract even more meaningful latent representations from the input images, allowing a more robust image-analysis pipeline. These can readily be incorporated into the work-flow, but have to be quality controlled through comparison with other networks and direct inspection of both the selected synthetic images and the discarded images.

The ALMA archive contains a vast collection of high-resolution images and data of young protostars. Combined with an increasing set of simulated protostellar systems, finding simulated analogues is a very valuable tool. It is our hope that the simulated analogues toolbox can ultimately accelerate the scientific learning rate from observations, by speeding up the analysis, allowing us to learn more about the dynamical state of the systems, and providing robust interpretations of the data.

\section*{Acknowledgements}

The research leading to these results has received funding from the Independent Research Fund Denmark through grant No.~DFF 8021-00350B (TH,RLK). This project has received funding from the European Union’s Horizon 2020 research and innovation program under the Marie Sklodowska-Curie grant agreement No. 847523 ‘INTERACTIONS’. The astrophysics HPC facility at the University of Copenhagen, supported by research grants from the Carlsberg, Novo, and Villum foundations, was used for carrying out the simulations, the analysis, and the long-term storage of the results. RLK also acknowledge funding from the Klaus Tschira Foundation. \texttt{yt} \citep{turk_yt:_2011}, \texttt{MATPLOTLIB} \citep{Hunter:2007} and \texttt{NumPy} \citep{harris2020array} were used to visualise and analyse these simulations. Variants of VGG16, VGG19 \citep{simonyan2014very}, InceptionV3 \citep{szegedy2016rethinking}, Resnet200 \citep{he2016identity}, Xception \citep{chollet2017xception}, DenseNet121 \citep{huang2017densely}, MobileNet \citep{Howard_2019_ICCV} were used to implement our algorithms using the packages of TensorFlow \citep{tensorflow} and Keras \citep{keras}. Special thanks to Evangelia Skoteinioti Stafyla for the illustration in \Cref{fig:cone}.

\section*{Data Availability}
All data and tools are available from the authors at reasonable request. 

\bibliographystyle{mnras}
\bibliography{references}
\bsp	
\label{lastpage}
\end{document}